\documentclass[amsmath,amssymb,aps,pre,reprint,showpacs]{revtex4-1}     
\usepackage{bm}
\usepackage{dcolumn}
\usepackage{graphicx}

\usepackage[caption=false]{subfig}
\usepackage{fancyref}
\begin{document}
\preprint{APS/123-QED}

\title{Emergence of stochastic dynamics in plane Couette flow}
\author{Rishabh Gvalani}
\affiliation{Department of Aeronautics, Imperial College London, London SW7 2AZ, UK}
\author{C\'edric Beaume}
 \email{c.m.l.beaume@leeds.ac.uk}
\affiliation{Department of Applied Mathematics, University of Leeds, Leeds LS2 9JT, UK}
\date{\today}

\begin{abstract}
Spatially localized states play an important role in transition to turbulence in shear flows (Kawahara, Uhlmann \& van Veen, {\it Annu. Rev. Fluid Mech.} {\bf 44}, 203 (2012)). Despite the fact that some of them are attractors on the separatrix between laminar and turbulent flows, little is known of their dynamics. We investigate here the temporal dynamics of such steady spatially localized solutions in the context of plane Couette flow. These solutions exist on oscillating branches in parameter space. We consider the saddle-nodes of these branches as initial conditions of simulations run with offset Reynolds numbers. We observe a relaminarization regime mostly characterized by deterministic dynamics and identify within this regime the existence of parameter intervals in which the results are stochastic and long-lived chaotic transients are observed. These results are obtained below the threshold for transition, shed light on the emergence of stochasticity in transitional plane Couette flow and will likely inform a range of shear flow configurations.
\pacs{PACS numbers}

\end{abstract}

\maketitle

Flow properties change dramatically as transition to turbulence occurs.
The flow becomes spatially and temporally chaotic and its interaction with bodies is largely affected: the drag behind vehicles is increased, the skin friction on duct walls is increased and the flow unsteadiness leads to a number of additional problems associated with material physics.
In most applications, this calls for adaptation or control to provide a viable process and maintain its efficiency.

Plane Couette flow is a popular configuration used to study this transition in which a three-dimensional viscous fluid is confined between two parallel walls moving in opposite directions and under no external forces.
This flow belongs to the family of subcritical shear flows for which the coexistence of the laminar state along with turbulent dynamics above a threshold Reynolds number suggests that there is a separatrix in phase space that one must cross to transition to turbulence.
Due to the nature of one of the bistable states, this separatrix is commonly referred to as the edge of chaos \cite{Skufca06,Schneider06} and hosts a number of exact solutions of prime dynamical importance \cite{Kawahara12}.
In small enough domains, the edge is structured by invariant self-sustained solutions such as the stationary Nagata solution \cite{Nagata90,Wang07,Schneider08}.
These solutions have a small number of unstable eigen-directions and are characterized by streamwise streaks, rolls and fluctuations that are kept in balance through a self-sustaining loop \cite{Waleffe97}.
For domains of large spanwise extent, the edge possesses attractors that take the form of spatially localized stationary or traveling wave states that can be thought of as bound states of the Nagata and laminar solutions \cite{Schneider10}.

Transition is generally studied for initial conditions generated by posttreated turbulent snapshots \cite{Avila11,Duguet11,Shi13}.
In this Letter, we take the complementary approach of considering exact spatially localized solutions as initial conditions.


We consider plane Couette flow in a doubly periodic domain.
The lengths are non-dimensionalised by half the separation between the walls $h$ such that the non-dimensional streamwise (resp. spanwise) extent of the domain reads $L_{x}$ (resp. $L_{z}$).
In the following, we set $L_{x} = 4 \pi$, following \cite{Schneider10}.
Velocities are non-dimensionalised by the speed of the walls $U$.
The no-slip boundary condition thus reads:
\begin{align}
\mathbf{u} (x,y=\pm 1,z)&=(\pm 1,0,0),
\end{align}
where ${\bf u}$ is the velocity field and $y=\pm1$ corresponds to the location of the walls.
This boundary condition is complemented with periodic boundary conditions in $x$ and $z$.
The flow is solved for using the Navier--Stokes equation
\begin{align}
\frac{\partial \mathbf{u}}{\partial t} + (\mathbf{u}.\nabla) \mathbf{u}&=-\nabla p +\frac{1}{Re}\nabla^{2}\mathbf{u}, 
\end{align}
along with the incompressibility condition
\begin{align}
\nabla \cdot \mathbf{u} = 0,
\end{align}
where $Re = Uh/\nu$ is the Reynolds number, $\nu$ is the fluid's dynamic viscosity, and $t$ and $p$ denote the nondimensional time and pressure. 

The linearly stable trivial laminar solution ${\bf U} = (y,0,0)$ is used to introduce the nontrivial velocity $\boldsymbol{u}=\mathbf{u}-\mathbf{U} = (u,v,w)$ where $u$ (resp. $v$, $w$) denotes its streamwise (resp. wall-normal, spanwise) component.
The equations and boundary conditions are invariant with respect to the reflection $\mathcal{R}: (u,v,w)(x,y,z) \longrightarrow -(u,v,w)(-x,-y,-z)$ and the shift-reflect symmetry: $\mathcal{S}: (u,v,w)(x,y,z) \longrightarrow (u,v,-w)(x+L_x/2,y,-z)$.
The imposition of the reflection symmetry yields stationary spatially localized edge-states while the imposition of the shift-reflect symmetry returns their traveling wave counterparts.
These two solutions live on branches that are intertwined in parameter space in a behavior known as snaking \cite{Schneider10,Gibson16}.
Snaking of localized states has been extensively studied in the context of the Swift--Hohenberg equation \cite{Burke06,Knobloch15} as well as in fluid dynamics \cite{Mercader11,Beaume13ddd1,Beaume13ddd2,Lojacono13}.
The decay of such states, however, has only been elucidated in the Swift--Hohenberg equation.
In particular, to the right of the snaking, the pattern gradually takes over the domain through nucleation events that occur at a temporal frequency proportional to the square root of the parametric distance to the right saddle-nodes of the snaking \cite{Burke06,Gandhi15,Knobloch15}.
As a result, perturbations in the direction of increasing Reynolds numbers are expected to lead to the increase of the localized solution width through the gradual motion of the fronts and nucleation of new rolls.
Lastly, the end state cannot be comprised of a periodic array of rolls, like the one in the Swift--Hohenberg equation: neither the Nagata, nor any other known exact solution, nor turbulence is stable below $Re \approx 325$ \cite{Shi13}.
For such values of the Reynolds number, a relaminarization mechanism eventually kicks in to annihilate any remaining nonlinear pattern.

\begin{figure*} 
\centering
\includegraphics[width=\linewidth]{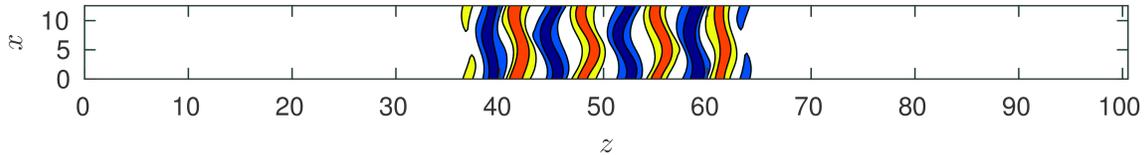}
\caption{Streamwise velocity $u(x,y=0,z)$ of the solution taken at $Re \approx 175.38 $ on the lowest right saddle-node of the snaking branch (see saddle-node $S_0$ in Figure \ref{snaking}). Blue (resp. red) indicates negative (resp. positive) values.}
\label{initsol}
\end{figure*}
\begin{figure} 
\includegraphics[scale=0.65]{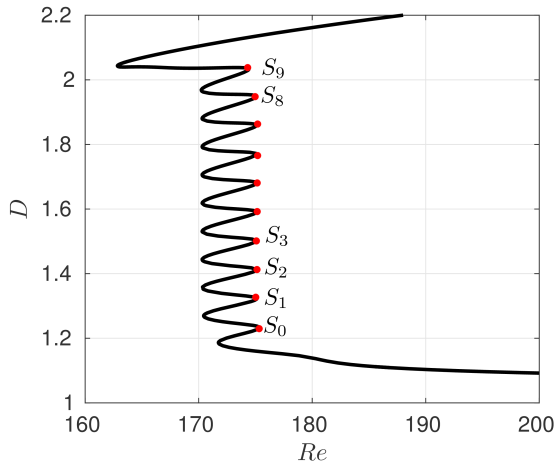}
\caption{Snaking of the equilibrium state of figure \ref{initsol} represented through the dissipation $D$ given by $V D=\int_{\Omega}|\nabla \times \mathbf{u}|^{2}\partial  \Omega$ as a function of the Reynolds number $Re$, where $V=2 L_x L_z$ is the volume of the domain $\Omega$.}
\label{snaking}
\end{figure}

We use the $\mathcal{R}$-symmetric localized edge state from \cite{Schneider10} at $Re = 400$, $L_x = 4\pi$ and $L_z = 16 \pi$ as a starting point for our investigation.
We first increase the domain size to $L_z = 32 \pi$ and continue the solution to lower values of the Reynolds number using a Newton--Raphson method to compute its snaking diagram.
The spatially localized exact solution is shown in Figure \ref{initsol} through its midplane ($y=0$) streamwise velocity at $Re \approx 175.38$.
It is comprised of dominant streamwise streaks (and rolls) and displays smaller amplitude oscillations in the streamwise direction imprinted by the fluctuations.
Other modes contribute but are substantially weaker \cite{Wang07}.
This solution exists down to $Re \approx 172$ and then oscillates in parameter space to form a snaking branch with saddle-node accumulation at $Re \approx 170$ and $Re_{sn} \approx 175$ (see Figure \ref{snaking}).
At the bottom saddle-nodes (Figure \ref{initsol}), the localized solution consists of $4$ pairs of counter rotating streamwise-oscillating rolls.
This pattern grows by one roll on each of its sides after each back and forth oscillation of the branch in $Re$.
It entirely fills the domain at the top of the branch.

The right saddle-nodes of the snaking are stored to serve as an initial condition for our simulations.
They are labeled: $S_{0},S_{1},S_{2}, \dots S_{9}$, with $S_{0}$ being the lowest saddle-node (see solution in Figure \ref{initsol}).
The numerical simulations are carried out using channelflow \cite{channelflow} with $N_x=32$ and $N_z=512$ Fourier points in the streamwise and spanwise directions and $N_y=33$ Chebyshev points in the wall-normal direction.
The initial conditions are perturbed in $Re$ and time-integrated until relaminarization.
We quantify the relaminarization time at any given $Re$ by $T^{*}$, defined as the time it takes for the initial condition to reach a small enough $L^{2}$-norm.
This condition is written:
\begin{equation}
||\boldsymbol{u}||=\sqrt{\frac{1}{V}\int_{\Omega}  \boldsymbol{u} \cdot \boldsymbol{u} \; \partial \Omega} \; < \, 0.1.
\end{equation}
Larger values than $0.1$ might not indicate relaminarization as nonlinear terms might still be important.
Lower values led to results that are polluted by the stable manifold of the laminar solution: as $Re$ is increased, the (negative) growth rate of least stable eigenmode of the laminar solution increases and it takes longer to approach the laminar state.
These observations are not related to the mechanism of selection of the end state.

\begin{figure} 
\centering
\includegraphics[width=7cm]{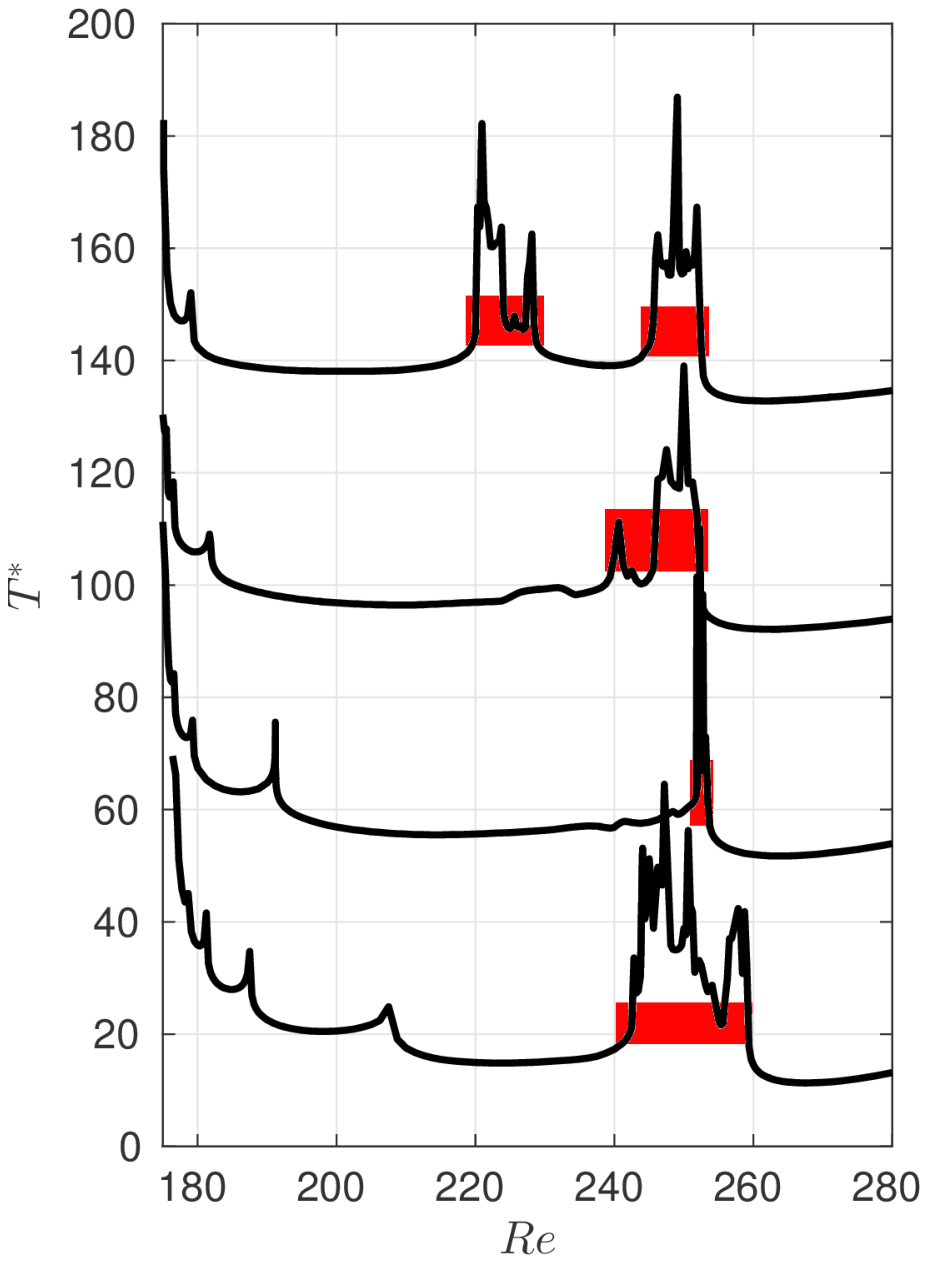}
\caption{Relaminarisation time $T^{*}$ as a function of the Reynolds number $Re$ for initial conditions $S_{i}$, $i=0, \dots, 3$. The results from saddle-nodes $S_{i}$, $i=1, \dots, 3$ have been offset in the following fashion $T^{*} \leftarrow T^{*} + 40 i$ to avoid superposition. Red regions indicate stochastic windows.}
\label{TvsRe}
\end{figure}

The results for the first $4$ saddle-nodes $S_i$, $i=0, \dots, 3$, are shown in Figure \ref{TvsRe}.
At the saddle-node, the initial condition is a fixed point and $T^* \rightarrow \infty$.
Away from it, the relaminarization time decays to $O(10)$ quantities.
This decay is not monotonic but rather displays plateaux interspersed with bursts.
For the $S_0$ initial condition, a number of these bursts/plateaux events occur between $Re \approx 175$ and $Re \approx 240$.
As $Re$ increases, the plateaux become progressively wider, the bursts less sharp and relaminarization occurs more rapidly.
The burst at $Re \approx 240$ gives rise to a window of stochastic relaminarization times highlighted in red in the figure which, in turn, gives rise to another plateau at $Re \approx 260$.
Lastly, past $Re \approx 285$, we observe (not shown) another window of stochastic results for which most initial conditions led to domain filling patterns.
Because these states are domain filling, finite size effects kick in and the subsequent dynamics are no longer relevant to the discussion in this Letter.
Similar observations are made for the other initial conditions, the differences being mainly quantitative, as exemplified in Figure \ref{TvsRe} for $S_1$, $S_2$ and $S_3$.
Overall, initial conditions with a larger pattern, i.e., taken further up the snaking, display similar dynamics but at values of the Reynolds number that are progressively shifted toward $Re_{sn}$.
This leads to a denser series of bursts in the vicinity of the saddle-nodes and to the earlier appearance of stochastic windows.

\begin{figure} 
\centering
\includegraphics[width=6cm]{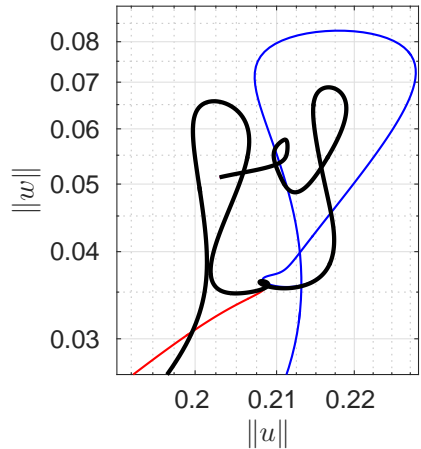}
\caption{Spanwise velocity norm $\|w\|$ as a function of the streamwise velocity norm $\|u\|$ for initial condition $S_{1}$ to the left of a burst (red, $Re \approx 191.1250$), close to its peak (black, $Re \approx 191.1934$) and to the right of the same burst (blue, $Re \approx 191.2031$).}
\label{jumplargecirc}
\end{figure}

In each of these simulations, relaminarization is triggered by the breaking of the self-sustaining loop.
Time-dependence in the form of oscillations in the fluctuation and roll amplitudes develops at the start of the simulations.
While these oscillations become larger and larger, the streaks only respond weakly, thus violating the self-sustaining balance.
The prominence of the above oscillatory behavior in the non-equilibrium dynamics is responsible for the bursts.
As $Re$ is increased past a burst, the trajectory of the solution in phase space produces one less loop before relaminarizing.
The bursts correspond to parameter values at which the solution trajectory follows a saddle: it is equally attracted by the relaminarizing manifold and the one that leads to another loop.
Figure \ref{jumplargecirc} shows three simulations illustrating this explanation: on the left of the burst (red) the solution relaminarizes directly, on the right (blue) it undergoes another loop in phase space while our closest simulation to the burst (black) shows an intermediate behavior.

\begin{figure} 
\centering
\includegraphics[width=8cm]{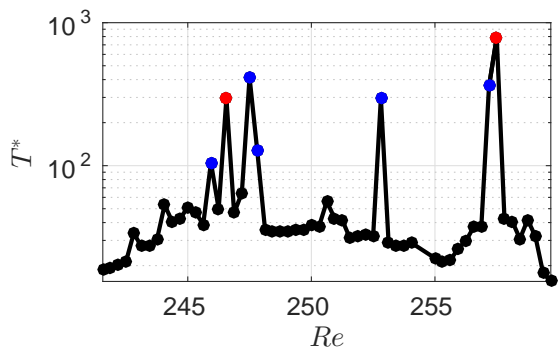}
\caption{Relaminarization time $T^{*}$ as a function of the Reynolds number $Re$ for saddle node $S_{0}$ and $Re \in [241.25;259.6875]$. The blue data points indicate simulations with $T^* > 100$ for which the localized pattern never grows to domain size. The red dots indicate simulations which transitioned to domain-filling chaos before decaying. The results are reported for a Reynolds number spacing of $0.3125$. Some of this data has been omitted in figure \ref{TvsRe} to avoid cluttering.}
\label{TvsRezoom}
\end{figure}

\begin{figure*} 
\centering
\includegraphics[width=15cm]{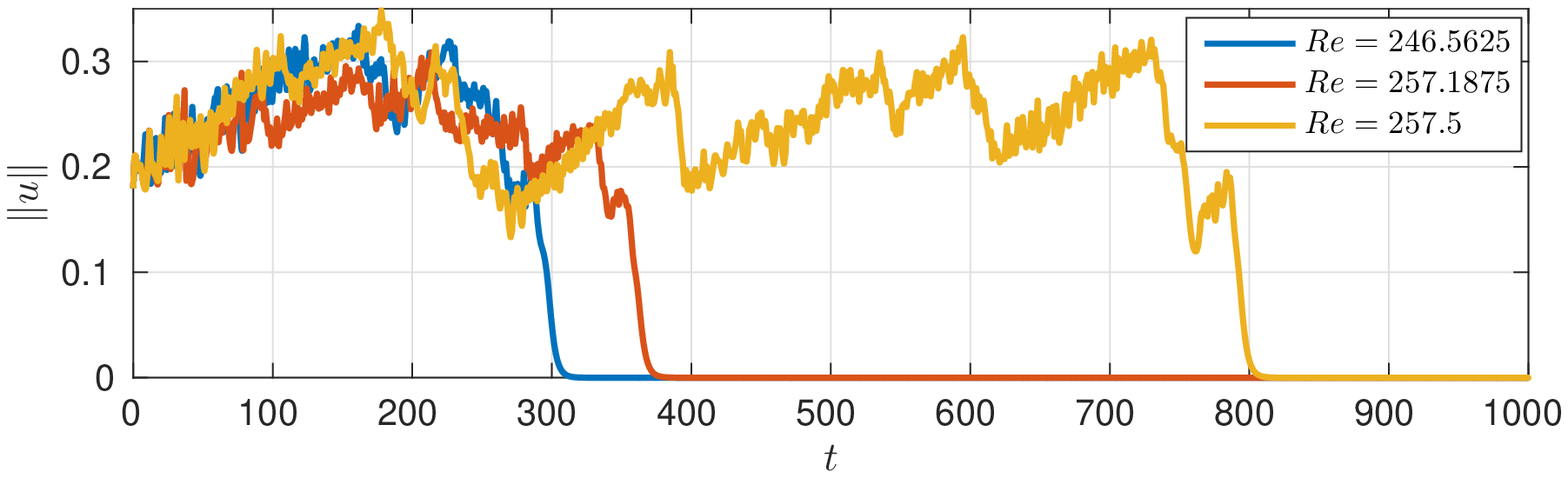}
\caption{Norm of the streamwise velocity $\|u\|$ as a function of time $t$ for three simulations yielding long-lasting chaos.}
\label{timeseries}
\end{figure*}

We now turn to the stochastic windows that appear after the initial series of bursts.
Figure \ref{TvsRe} shows that stochastic windows are initiated by a large and sudden increase in the relaminarization time.
Within these windows, the relaminarization time is increased compared to the surrounding plateaux to values that are typically twice larger and displays strong variations with $Re$.
Figure \ref{TvsRezoom} shows the relaminarization times in a logarithmic plot for the stochastic window associated with $S_0$.
They mostly take values between $T^*=20$ and $T^*=60$, while the surrounding plateaux hardly reach $T^* = 15$.
The variations of $T^*$ become stochastic and can lead to large relaminarization times.
Seven of the $57$ simulations (12$\%$) run within this window reached values above $T^*=100$, with the longest relaminarization time recorded at $T^* \approx 793$ for $Re = 257.5$.
These long simulations display chaotic dynamics during which the the localized pattern spreads in the spanwise direction through irregular nucleation events adding one roll to the pattern.
During their evolution, the rolls mostly keep their global structure but their amplitude and fluctuations vary greatly.
The pattern behaves like a conglomerate of clusters, each comprised of synchronized adjacent rolls.
Phase jumps between adjacent clusters are observed.
Our simulations have revealed only one decay mechanism: in a given cluster, the fluctuations become so weak that the rolls straighten and start diffusing away.
This mechanism does not act on an isolated roll but on an entire cluster and can create instantaneous localized states comprised of two separate pulses.
We exemplify the dynamics of this such long-lasting chaotic transients by three simulations in Figure \ref{timeseries}.
Two of these simulations involve domain-filling states (red dots in Figure \ref{TvsRezoom}): $Re=246.5625$ and $Re = 257.5$ which relaminarize at $T^* \approx 296$ and $T^* \approx 793$ respectively.
The pattern of the simulation run at $Re = 257.1875$ never filled more than $80\%$ of the domain and the simulation relaminarized at $T^* \approx 360$.
During these simulations, the norm of the streamwise velocity is seen to increase for long periods before undergoing abrupt events where it decreases.
The former dynamics correspond to the growth of the pattern via nucleation events while the latter ones correspond to cluster annihilation.



In this letter, we have investigated the dynamics of stationary spatially localized exact solutions of plane Couette flow when perturbed outside equilibrium.
These states live on snaking branches in the region $170 < Re < 175$ and depinning theory predicts that for $175 < Re$, new rolls will be gradually added to the structure making it grow in the spanwise direction \cite{Burke06,Gandhi15,Knobloch15}.
Earlier work observed depinning as a competing mechanism with decay in the vicinity of the snaking and survival of the localized state up to $5000$ time units \cite{Duguet11}.
In our case, we observe faster relaminarization: no localized state survived more than $1000$ time units.
We believe the different parameter values to be responsible for departures in the relaminarization times.
In addition, our choice of initial condition has allowed us to shed light on the competition between stochastic and deterministic dynamics in transitional plane Couette flow, in particular through the identification of stochastic windows.
A detailed comparison with different domain sizes and initial conditions is in preparation and will be reported elsewhere.

Our initial conditions are taken at $Re \approx 175$ and are allowed to evolve for different values of the Reynolds number.
As the Reynolds number is increased, the relaminarization time decreases through a succession of plateaux interspersed with bursts, which are shown to be related to oscillatory dynamics.
Our simulations have unravelled the existence of stochastic windows characterized by a substantial increase in the  system's sensitivity to the initial condition and in the duration of the chaotic transient.
These stochastic windows complement observations in \cite{Duguet11} that deterministic behavior competes with stochasticity.
These results are robust: they are qualitatively similar for all the initial conditions tested and we believe that they are applicable to a range of subcritical flow configurations.


The authors acknowledge J. Gibson and T. Schneider for providing some exact solutions and M. Salewski and J. Gibson for software support.

\bibliography{oscilPRL}

\end{document}